# Alternative perspective on photonic tunneling

**Zhi-Yong Wang[1*], Cai-Dong Xiong[1], Bing He[2]**

[1]*School of Physical Electronics, University of Electronic Science and Technology of China, Chengdu 610054, CHINA*

[2]*Department of Physics and Astronomy, Hunter College of the City University of New York, 695 Park Avenue, New York, NY 10021, USA*

A relativistic quantum-mechanical description of guided waves is given, based on which we present an alternative way to describe and interpret the propagation of electromagnetic wave packets through an undersized waveguide. In particular, we show that the superluminal phenomenon of evanescent modes is actually a known conclusion in quantum field theory, and it preserves a quantum-mechanical causality.

## I. INTRODUCTION

Traditionally, the propagation of electromagnetic wave packets through an undersized waveguide is interpreted in terms of "photonic tunneling", which is based on a mathematical analogy between the Helmholtz equation describing evanescent modes and the nonrelativistic Schrödinger equation describing a quantum-mechanical tunneling [1-3]. However, this theoretical framework has the following shortcomings: firstly, the photon's equation of motion is a relativistic one while the Schrödinger equation is a nonrelativistic one, and then this analogy is reasonable in mathematics but not in physics; secondly, an appropriate description for the propagation of evanescent modes should be based on the photon's quantum mechanics itself, rather than a quantum-mechanical analogy.

On the other hand, in QED the elementary field quantity is a four-dimensional (4D) electromagnetic potential, the quantum theory of the photon is directly a quantum field theory, while the photon's relativistic quantum mechanics (i.e., the first-quantized theory) is absent. However, in spite of QED's great success as well as the traditional conclusion that single photon cannot be localized [4-5], there have been many attempts to develop photon wave mechanics which is based on the concept of photon wave function and contains the first-quantized theory of the photon [6-13], and some recent studies have shown that photons can be localized in space [14-16]. These efforts have both theoretical and practical interests. To present a rigorous argument for the existence of evanescent modes' superluminal behavior, one can resort to photon wave mechanics in which photon wave function is formed by electromagnetic field intensities and can be regarded as energy-density amplitude (being an extension for the concept of probability amplitude in nonrelativistic quantum mechanics). Choosing a different photon wave function (i.e., a different representation of the Lorentz-group), one can obtain a different description of photon wave mechanics, but all of them can give the same physical conclusions about the existence of superluminal behaviors. However, we would rather choose a 6×1 photon wave function that transforms according to the $(1,0)\oplus(0,1)$ spinor representation of the Lorentz group, which due to the following reasons: 1) in contrast to the photon wave mechanics developed in previous literatures, the Dirac-like equation in our formalism has the Lorentz-covariant form, and there is a closely analogy between its mathematical structures and those of the Dirac equation, which is an aid to our discussions. In particular, via photon wave mechanics developed in our formalism, a relativistic quantum theory of photons inside a waveguide, at both the first- and second-quantized levels, can be developed. 2) The previous photon wave mechanics is short of a canonically theoretical framework from the

first- to second-quantized levels, where the negative-energy solution and the solution describing the admixture of the longitudinal and scalar photon are usually discarded by hand, these will present some problems. In contrast to which, in our formalism, photon wave mechanics has a systematically framework at both levels (first and second) of quantization, and here the negative-energy solution is reasonably preserved (this is in agreement with the standard relativistic quantum theory); the solution describing the admixture of the longitudinal and scalar photon states is kept and physically its contribution to energy and momentum is discarded in a natural way.

In view of all mentioned above, in this paper we will try to provide a relativistic quantum-mechanical description for guided waves, and at its second-quantized level, we present an alternative way to describe and interpret the propagation of electromagnetic wave packets in undersized waveguides. In addition to our work presented here, recently QED-based studies of evanescent modes came to the same conclusion, and a successful test of these predictions with experimental data has been presented [17]. In the following, the natural units of measurement ($\hbar = c = 1$) is applied, repeated indices must be summed according to the Einstein rule, and the space-time metric tensor is chosen as $g^{\mu\nu} = \mathrm{diag}(1, -1, -1, -1)$, $\mu, \nu = 0, 1, 2, 3$. For our convenience, let $x^{\mu} = (t, -\boldsymbol{x})$, instead of $x^{\mu} = (t, \boldsymbol{x})$, denote the contravariant position 4-vector (and so on), and then in our case $\hat{p}_{\mu} = \mathrm{i}\,\partial/\partial x^{\mu} \equiv \mathrm{i}\partial_{\mu} = \mathrm{i}(\partial_t, -\nabla)$ denote 4D momentum operators.

## II. PHOTON FIELD AS A SIX-COMPONENT SPINOR FIELD

In vacuum the electric field, $\boldsymbol{E} = (E_1, E_2, E_3)$, and the magnetic field, $\boldsymbol{B} = (B_1, B_2, B_3)$, satisfy the Maxwell equations

$$\nabla\times\boldsymbol{E}=-\partial_t\boldsymbol{B},\ \ \nabla\times\boldsymbol{B}=\partial_t\boldsymbol{E}, \tag{1}$$

$$\nabla\cdot\boldsymbol{E}=0,\ \ \nabla\cdot\boldsymbol{B}=0. \tag{2}$$

Let $(\tau_i)_{jk}=-\mathrm{i}\varepsilon_{ijk}$ , $i,j,k=1,2,3$ , where $\varepsilon_{ijk}$ is a totally antisymmetric tensor with $\varepsilon_{123}=1$. By means of matrix vector $\boldsymbol{\tau}=(\tau_1,\tau_2,\tau_3)$ and the quantities ($I_{n\times n}$ denotes the $n\times n$ unit matrix, $n=2,3,4...$)

$$\beta_0=\begin{pmatrix} I_{3\times3} & 0 \\ 0 & -I_{3\times3}\end{pmatrix},\ \boldsymbol{\beta}=\begin{pmatrix} 0 & \boldsymbol{\tau} \\ -\boldsymbol{\tau} & 0\end{pmatrix}, \tag{3}$$

$$E=\begin{pmatrix} E_1 \\ E_2 \\ E_3\end{pmatrix},\ B=\begin{pmatrix} B_1 \\ B_2 \\ B_3\end{pmatrix}, \tag{4}$$

$$\psi=\frac{1}{\sqrt{2}}\begin{pmatrix} E \\ \mathrm{i}B\end{pmatrix}, \tag{5}$$

one can rewrite the Maxwell equations as a *Dirac-like equation*

$$\mathrm{i}\beta^{\mu}\partial_{\mu}\psi(x)=0,\ \text{or}\ \ \mathrm{i}\partial_t\psi(x)=\hat{H}\psi(x), \tag{6}$$

where $\hat{H}=-\mathrm{i}\boldsymbol{\chi}\cdot\nabla$ represents the Hamiltonian of the free photon and $\boldsymbol{\chi}=\beta_0\boldsymbol{\beta}$ . Let $\hat{\boldsymbol{L}}=\boldsymbol{x}\times(-\mathrm{i}\nabla)$ be the orbital angular momentum operator, one can easily obtain $[\hat{H},\hat{\boldsymbol{L}}+\boldsymbol{S}]=0$ , where $\boldsymbol{S}=I_{2\times2}\otimes\boldsymbol{\tau}$ satisfying $\boldsymbol{S}\cdot\boldsymbol{S}=1(1+1)I_{6\times6}$ represents the spin matrix of the spin-1 field. In fact, one can show that the 6×1 spinor $\psi(x)$ transforms according to the $(1,0)\oplus(0,1)$ representation of the Lorentz group, and can prove that $(\beta^{\mu}\partial_{\mu})(\beta_{\nu}\partial^{\nu})=\partial^{\mu}\partial_{\mu}+\varOmega$, where $\varOmega\psi(x)=0$ is identical with the transverse conditions given by Eq. (2), thus Eq. (6) implies that the wave equation $\partial^{\mu}\partial_{\mu}\psi(x)=0$, where

$$\Omega=I_{2\times2}\otimes[\begin{pmatrix}\nabla_1 \\ \nabla_2 \\ \nabla_3\end{pmatrix}\begin{pmatrix}\nabla_1 & \nabla_2 & \nabla_3\end{pmatrix}], \tag{7}$$

Let $k_{\mu}=(\omega,\boldsymbol{k})$ denote the 4D momentum of photons ($\hbar=c=1$), where $\omega$ is the frequency and $\boldsymbol{k}$ the wave-number vector. The fundamental solutions of the Dirac-like equation (6) are represented by the positive- and negative-frequency components, respectively

$$\begin{cases}\phi_{k,\lambda}(x) = (\omega/V)^{1/2} f(k,\lambda)\exp(-\mathrm{i}k\cdot x)\\ \phi_{-k,\lambda}(x) = (\omega/V)^{1/2} g(k,\lambda)\exp(\mathrm{i}k\cdot x)\end{cases}, \tag{8}$$

where $\int (1/V)\mathrm{d}^3x = 1$, $\lambda = \pm 1, 0$, and

$$f(k,\lambda) = \frac{1}{\sqrt{1+\lambda^2}}\begin{pmatrix}\varepsilon(\boldsymbol{k},\lambda)\\ \lambda\varepsilon(\boldsymbol{k},\lambda)\end{pmatrix}, \tag{9}$$

$$g(k,\lambda) = \frac{1}{\sqrt{1+\lambda^2}}\begin{pmatrix}\lambda\varepsilon(\boldsymbol{k},\lambda)\\ \varepsilon(\boldsymbol{k},\lambda)\end{pmatrix}, \tag{10}$$

where $\varepsilon(\boldsymbol{k},0)$ is the longitudinal polarization vector of the photons, while $\varepsilon(\boldsymbol{k},1)$ and $\varepsilon(\boldsymbol{k},-1)$ are the right- and left-hand circular polarization vectors, respectively. Let $\varepsilon^*(\boldsymbol{k},-1)$ denote the complex conjugate of $\varepsilon(\boldsymbol{k},-1)$ (and so on), in matrix form,

$$\varepsilon(\boldsymbol{k},1) = \varepsilon^*(\boldsymbol{k},-1) = \frac{1}{\sqrt{2}|\boldsymbol{k}|}\begin{pmatrix}\dfrac{k_1k_3 - \mathrm{i}k_2|\boldsymbol{k}|}{k_1 - \mathrm{i}k_2}\\ \dfrac{k_2k_3 + \mathrm{i}k_1|\boldsymbol{k}|}{k_1 - \mathrm{i}k_2}\\ -(k_1+\mathrm{i}k_2)\end{pmatrix},\quad \varepsilon(\boldsymbol{k},0) = \frac{1}{|\boldsymbol{k}|}\begin{pmatrix}k_1\\ k_2\\ k_3\end{pmatrix}. \tag{11}$$

Correspondingly, $\lambda = \pm 1, 0$ represent the spin projections in the direction of $\boldsymbol{k}$. As we know, when the electromagnetic field are described by an 4D electromagnetic potential, there involves four polarization vectors respectively describing four kinds of photons, while described by the 6×1 spinor $\psi(x)$ (constructed by the electromagnetic field intensities), there only involves three polarization vectors. Therefore, in our framework, the $\lambda = 0$ solution describes the admixture of the longitudinal and scalar photon states, while the $\lambda = \pm 1$ solutions describe the transverse photon states.

Let $\varepsilon^+$ stand for the Hermitian conjugate of $\varepsilon$ (and so on), one has

$$\begin{cases} \varepsilon^+(\boldsymbol{k},\lambda)\varepsilon(\boldsymbol{k},\lambda') = \delta_{\lambda\lambda'} \\ \sum_{\lambda}\varepsilon(\boldsymbol{k},\lambda)\varepsilon^+(\boldsymbol{k},\lambda) = I_{3\times 3} \end{cases}. \tag{12}$$

The fundamental solutions given by Eq. (8) satisfy the orthonormality and completeness relations

$$\begin{cases} \int \phi^+_{k,\lambda}\phi_{k',\lambda'}\mathrm{d}^3x = \int \phi^+_{-k,\lambda}\phi_{-k',\lambda'}\mathrm{d}^3x = \omega\delta_{\lambda\lambda'}\delta_{\boldsymbol{kk}'} \\ \int \phi^+_{k,\lambda}\phi_{-k'_0,\boldsymbol{k}',\lambda'}\mathrm{d}^3x = \int \phi^+_{-k_0,\boldsymbol{k},\lambda}\phi_{k',\lambda'}\mathrm{d}^3x = 0 \end{cases}, \tag{13}$$

$$\sum_{\lambda}\int (\phi_{k,\lambda}\phi^+_{k,\lambda} + \phi_{-k_0,\boldsymbol{k},\lambda}\phi^+_{-k_0,\boldsymbol{k},\lambda})\mathrm{d}^3x = \omega I_{6\times 6}\ . \tag{14}$$

Substituting $\varphi(k)\exp(-\mathrm{i}k^\mu x_\mu)$ into Eq. (6) one has $\det(\omega - \boldsymbol{\chi}\cdot\boldsymbol{k}) = 0$. Let $k^\mu_{\mathrm{T}} = (\omega_{\mathrm{T}}, -\boldsymbol{k}_{\mathrm{T}})$ denote the 4D momentum of the transverse photons (corresponding to the $\lambda = \pm 1$ solutions), while $k^\mu_{\mathrm{L}} = (\omega_{\mathrm{L}}, -\boldsymbol{k}_{\mathrm{L}})$ the 4D momentum of the longitudinal and scalar photons (both corresponding to the $\lambda = 0$ solution), using $\det(\omega - \boldsymbol{\chi}\cdot\boldsymbol{k}) = 0$ and Eq. (2) one has

$$\omega_{\pm 1} \equiv \omega_{\mathrm{T}} = |\boldsymbol{k}_{\mathrm{T}}|,\ \ \omega_0 \equiv \omega_{\mathrm{L}} = |\boldsymbol{k}_{\mathrm{L}}| = 0\,, \tag{15}$$

which in agreement with the traditional conclusions that the contributions of the longitudinal and scalar photons to the energy and momentum cancel each other. In the following treatment, for convenience the $\lambda = 0$ solution is kept and only in the end let $k^\mu_L \to 0$ for the $\lambda = 0$ solution.

Consider that antiphotons are identical with photons, we expand the general solution of Eq. (6) via the fundamental solutions $\phi_{k,\lambda}$ and $\phi_{-k,\lambda}$ as

$$\psi(x) = \frac{1}{\sqrt{2}}\sum_{\boldsymbol{k},\lambda}[a(k,\lambda)\phi_{k,\lambda} + a^+(k,\lambda)\phi_{-k,\lambda}]\,. \tag{16}$$

Let $\bar{\psi} \equiv \psi^+\beta^0$ (and so on), consider that $\psi(x)$ corresponds to the $(1,0)\oplus(0,1)$ representation of the Lorentz group, one can prove that

$$\mathcal{L} = \bar{\psi}(x)(\mathrm{i}\beta^{\mu}\partial_{\mu})\psi(x) \tag{17}$$

is a Lorentz scalar, but its dimension is $[1/\text{length}]^5$ rather than $[1/\text{length}]^4$, and then we call it pseudo-Lagrangian density. Applying the variational principle in $A = \int \mathcal{L}\mathrm{d}^4x$ one can obtain the Dirac-like equation (6). Consider that the frequency $\omega > 0$, we can apply the inverse of the operator $-\mathrm{i}\partial_t$, denoted as $(-\mathrm{i}\partial_t)^{-1}$, to define the canonical momentum conjugating to $\psi(x)$ as

$$\pi \equiv (-\mathrm{i}\frac{\partial}{\partial t})^{-1}\frac{\partial \mathcal{L}}{\partial \dot{\psi}} = \frac{\partial \mathcal{L}'}{\partial \dot{\psi}} = (-\mathrm{i}\partial_t)^{-1}\mathrm{i}\,\psi^{+}, \tag{18}$$

where $\dot{\psi} = \partial_t\psi$, and

$$\mathcal{L}' \equiv [(-\mathrm{i}\partial_t)^{-1}\bar{\psi}(x)]\mathrm{i}\beta^{\mu}\partial_{\mu}\psi(x) \tag{19}$$

is called pseudo-Lagrangian density with dimension $[1/\text{length}]^4$. The conserved charges related to the invariance of $\mathcal{L}'$ under the space-time translations, are the 4D momentum (say, $p^{\mu} = (H, \boldsymbol{p})$) of the photon field, and they can be expressed as

$$H = \int [\pi(x)\dot{\psi}(x) - \mathcal{L}']\mathrm{d}^3x, \tag{20}$$

$$\boldsymbol{p} = -\int [\pi(x)\nabla\psi(x)]\mathrm{d}^3x. \tag{21}$$

To obtain the canonical commutation relations in our formalism, we must start from the traditional ones satisfied by the 4D electromagnetic potential (say, $A^{\mu}(x)$, $\mu = 0,1,2,3$)

$$[A^{\mu}(x), A^{\nu}(y)] = -\mathrm{i}g^{\mu\nu}D(x-y), \tag{22}$$

where

$$\mathrm{i}D(x) \equiv \int \frac{\mathrm{d}^3k}{(2\pi)^3}\frac{1}{2\omega}[\exp(-\mathrm{i}k\cdot x) - \exp(\mathrm{i}k\cdot x)]. \tag{23}$$

Using Eqs. (4)-(5), (22)-(23) and

$$E^{i} = \partial^{i}A^{0} - \partial^{0}A^{i},\ \ B^{i} = \varepsilon^{ijk}\partial_{j}A_{k},\quad i,j,k = 1,2,3, \tag{24}$$

one can obtain the following canonical commutation relations

$$\begin{cases}[\psi_i(\boldsymbol{x},t),\pi_j(\boldsymbol{x}',t)]=-\mathrm{i}\delta_{\mathrm{T}ij}\delta^3(\boldsymbol{x}-\boldsymbol{x}')/2\\ [\psi_{i+3}(\boldsymbol{x},t),\pi_{j+3}(\boldsymbol{x}',t)]=-\mathrm{i}\delta_{\mathrm{T}ij}\delta^3(\boldsymbol{x}-\boldsymbol{x}')/2\end{cases}, \tag{25}$$

with the other commutators vanishing, where $\delta_{\mathrm{T}ij}\equiv\delta_{ij}-(\partial_i\partial_j/\nabla^2)$ is the transverse delta function. Using Eqs. (16), (18) and (25), we get the following commutation relations

$$[a(k,\lambda),a^+(k',\lambda')]=\delta_{\boldsymbol{kk}'}\delta_{\lambda\lambda'}, \tag{26}$$

with other commutators vanishing. For the moment, Eq. (6) can be obtained from Heisenberg's equation of motion $\partial_t\psi=\mathrm{i}[H,\psi]$. Using $k_{\mathrm{L}}^{\mu}=(\omega_{\mathrm{L}},-\boldsymbol{k}_{\mathrm{L}})\to 0$ and Eqs. (20), (21), (26), one has

$$H=\sum_{\boldsymbol{k}}\sum_{\lambda=\pm1}\omega[a^+(k,\lambda)a(k,\lambda)+(1/2)], \tag{27}$$

$$\boldsymbol{p}=\sum_{\boldsymbol{k}}\sum_{\lambda=\pm1}\boldsymbol{k}[a^+(k,\lambda)a(k,\lambda)]. \tag{28}$$

This is in agreement with the traditional theory.

Furthermore, applying Eqs. (9), (10) and (12) one can prove that

$$\sum_{\lambda=\pm1}[f(k,\lambda)\bar{f}(k,\lambda)]=\sum_{\lambda=\pm1}[g(k,\lambda)\bar{g}(k,\lambda)]=(\beta_\mu k^\mu/2\omega)(I_{6\times6})_\perp, \tag{29}$$

where

$$(I_{6\times6})_\perp\equiv I_{2\times2}\otimes\sum_{\lambda=\pm1}\varepsilon(\boldsymbol{k},\lambda)\varepsilon^+(\boldsymbol{k},\lambda). \tag{30}$$

In our formalism, let us define the free Feynman propagator of the photon as:

$$\mathrm{i}R_{\mathrm{f}}(x_1-x_2)\equiv(\mathrm{i}\partial/\partial t_1)^{-1}\langle 0|\mathrm{T}\psi(x_1)\bar{\psi}(x_2)|0\rangle, \tag{31}$$

where $|0\rangle$ stands for the vacuum state, T is the time order symbol, i.e.

$$\mathrm{T}\varphi_\alpha(x_1)\bar{\psi}_\beta(x_2)=\begin{cases}\varphi_\alpha(x_1)\bar{\psi}_\beta(x_2),\ t_1>t_2\\ \bar{\psi}_\beta(x_2)\varphi_\alpha(x_1),\ t_2>t_1\end{cases}. \tag{32}$$

Applying $k_{\mathrm{L}}^{\mu}\to 0$ and Eq. (29) one can prove that

$$\mathrm{i}R_{\mathrm{f}}(x_1-x_2)=\mathrm{i}\beta^\mu\partial_\mu\delta_{\mathrm{T}}\mathrm{i}\Delta(x_1-x_2). \tag{33}$$

where $\partial_\mu = \partial/\partial x_1^\mu$ , $\delta_{\mathrm{T}}$ is the transverse delta function: $\delta_{\mathrm{T}ij} \equiv \delta_{ij} - (\partial_i \partial_j / \nabla^2)$ , $i, j = 1,2,3$ , $\mathrm{i}\Delta(x)$ is the free Feynman propagator of massless scalar fields:

$$\mathrm{i}\Delta(x) = \int_{-\infty}^{+\infty} \frac{\mathrm{d}^4 k}{(2\pi)^4} \frac{\mathrm{i}}{k^2 + \mathrm{i}\rho} \exp(-\mathrm{i}k \cdot x) , \tag{34}$$

where $\rho$ is an infinitesimal real quantity. One can examine that

$$(\mathrm{i}\beta^\mu \partial_\mu) R_{\mathrm{f}} (x_1 - x_2) = \delta^4 (x_1 - x_2) , \tag{35}$$

That is, $\mathrm{i}R_{\mathrm{f}}(x_1 - x_2)$ is the Green function of the Dirac-like equation (6).

It is very important to note that, there is an alternative way of developing our theory: to define an inner product and the mean value of an operator $\hat{L}$ as $\langle \psi | \psi' \rangle \equiv \int \psi^+ \psi' \mathrm{d}^3 x$ and $\langle L \rangle \equiv \int \psi^+ \hat{L} \psi \mathrm{d}^3 x$, respectively, the solutions of the Dirac-like equation (6) can be rechosen as those having the dimension of $[1/\text{length}]^{3/2}$ (rather than $[1/\text{length}]^2$), such that they no longer correspond to the $(1,0) \oplus (0,1)$ representation of the Lorentz group, but rather to particle-number density amplitude. This presents no problem provided that one can ultimately obtain Lorentz-covariant observables. In fact, in nonrelativistic quantum mechanics, the wave functions of all kinds of particles stand for probability amplitudes with the dimension of $[1/\text{length}]^{3/2}$, and do not correspond to any representation of the Lorentz group. In the first-quantized sense, the field quantities presented in our formalism can be regarded as charge-density amplitude or particle-number density amplitude, which is an extension for the concept of probability amplitude. For the moment, the normalization factor presented in $\psi(x)$ is chosen as $1/\sqrt{V}$ instead of the original $\sqrt{\omega}/\sqrt{V}$ (see Eq. (8)), and the factor of $\omega$ no longer appears on the right of Eq. (13) and (14). In particular, now the Lagrangian density $\mathcal{L} = \bar{\psi}(x)(\mathrm{i}\beta^\mu \partial_\mu)\psi(x)$ has the dimension of $[1/\text{length}]^4$, the canonical momentum conjugating to $\psi(x)$ is defined according to the traditional form of $\pi = \partial \mathcal{L}/\partial \dot{\psi} = \mathrm{i}\psi^{+}$ , and the free Feynman propagator of the photon is redefined as $\mathrm{i}R_{\mathrm{f}}(x_1 - x_2) \equiv \langle 0 | \mathrm{T}\, \psi(x_1)\bar{\psi}(x_2) | 0 \rangle$. However, all the final physical conclusions are preserved

By the way, the Dirac-like equation (6) is valid for all kinds of electromagnetic fields outside a source, including the time-varying and static fields, the transverse and longitudinal fields, and those generated by an electrical or magnetic multipole moment, etc. In particular, Eqs. (8)-(10) and (5) together show that the symmetry between the positive- and negative-energy solutions corresponds to the duality between the electric and magnetic fields, rather than to the usual particle-antiparticle symmetry. Moreover, in the positive-energy solutions $E$ has longitudinal component while $B$ not; in the negative-energy solutions $B$ has longitudinal component while $E$ not. Therefore, if the positive-energy solutions are regarded as the fields produced by an electric $N$-pole moment, then the negative-energy solutions can be regarded as the fields produced by a magnetic $N$-pole moment ($N$=2, 4, 6,…) [18].

## III. RELATIVISTIC QUANTUM THEORY OF GUIDED WAVES

As we know, on one hand, the TEM mode (both the electric and magnetic fields perpendicular to the direction of propagation) cannot propagate in a single conductor transmission line, only those modes in the form of transverse electric (TE) and transverse magnetic (TM) modes can propagate in the waveguide. On the other hand, however, there has another useful way of looking at the propagation, that is, the guided waves can be viewed as the superposition of two sets of plane waves (i.e., the TEM waves) being continually reflected back and forth between perfectly conducting walls and zigzagging down the waveguide, the two sets of plane waves have the same amplitudes and frequencies, but reverse phases. Usually, the propagation of the electromagnetic wave through an ideal and uniform waveguide is described by a wave equation in (1+1) D space-time. However, in our formalism, we will generally place the waveguide along an arbitrary 3D spatial direction, by which we will show that guided waves have the same behaviors as de Broglie matter waves, and in terms of the 6×1 spinor defined by Eq. (5), we obtain a relativistic quantum description for the guided waves at both levels (first and second) of quantization.

In a Cartesian coordinate system spanned by an orthonormal basis $\{\boldsymbol{e}_1,\boldsymbol{e}_2,\boldsymbol{e}_3\}$ with $\boldsymbol{e}_3=\boldsymbol{e}_1\times\boldsymbol{e}_2$, we assume that a hollow metallic waveguide is placed along the direction of $\boldsymbol{e}_3$, and the waveguide is a straight rectangular pipe with the transversal dimensions $b_1$ and $b_2$, let $b_1>b_2$ without loss of generality. It is also assumed that the waveguide is infinitely long and its conductivity is infinite, and the electromagnetic source is localized at infinity. In the Cartesian coordinate system $\{\boldsymbol{e}_1,\boldsymbol{e}_2,\boldsymbol{e}_3\}$, let $k_\mu=(\omega,\boldsymbol{k})$ denote the 4D momentum of photons inside the waveguide, then the wave-number vector is $\boldsymbol{k}=\sum_i\boldsymbol{e}_i k_i=(k_1,k_2,k_3)$ and the frequency $\omega=|\boldsymbol{k}|$, where $k_1=r\pi/b_1$ and $k_2=s\pi/b_2$ ($r=1,2,3...$, $s=0,1,2...$), and the cutoff frequency of the waveguide is $\omega_{crs}=\sqrt{k_1^2+k_2^2}=\pi\sqrt{(r/b_1)^2+(s/b_2)^2}$. For simplicity, we shall restrict our discussion to the lowest-order cutoff frequency $\omega_{\mathrm{c}}=\pi/b_1$.

We define the effective mass of photons inside the waveguide as $m=\omega_{\mathrm{c}}$, then the photon energy $E=\omega$ satisfies $E^2=k_3^2+m^2$. To obtain a Lorentz covariant formulation, let us rechoose a Cartesian coordinate system formed by an orthonormal basis $\{\boldsymbol{a}_1,\boldsymbol{a}_2,\boldsymbol{a}_3\}$ with $\boldsymbol{a}_3=\boldsymbol{a}_1\times\boldsymbol{a}_2$, such that in the new coordinate system, one has $\boldsymbol{e}_3k_3=\sum_j\boldsymbol{a}_j p_j=\boldsymbol{p}$. That is, in the new coordinate system, the waveguide is put along an arbitrary 3D spatial direction. Let $k_3\geq 0$ without loss of generality. If $k_3=|\boldsymbol{p}|>0$, in the coordinate system $\{\boldsymbol{a}_1,\boldsymbol{a}_2,\boldsymbol{a}_3\}$, the unit vectors $\boldsymbol{e}_i$ ($i=1,2,3$) can be expressed as

$$\begin{cases} \boldsymbol{e}_1 = \boldsymbol{e}(\boldsymbol{p},1) = (\dfrac{p_1^2 p_3 + p_2^2 |\boldsymbol{p}|}{|\boldsymbol{p}|(p_1^2+p_2^2)}, \dfrac{p_1 p_2 p_3 - p_1 p_2 |\boldsymbol{p}|}{|\boldsymbol{p}|(p_1^2+p_2^2)}, -\dfrac{p_1}{|\boldsymbol{p}|}) \\ \boldsymbol{e}_2 = \boldsymbol{e}(\boldsymbol{p},2) = (\dfrac{p_1 p_2 p_3 - p_1 p_2 |\boldsymbol{p}|}{|\boldsymbol{p}|(p_1^2+p_2^2)}, \dfrac{p_2^2 p_3 + p_1^2 |\boldsymbol{p}|}{|\boldsymbol{p}|(p_1^2+p_2^2)}, -\dfrac{p_2}{|\boldsymbol{p}|}) \\ \boldsymbol{e}_3 = \boldsymbol{e}(\boldsymbol{p},3) = \boldsymbol{e}(\boldsymbol{p},1) \times \boldsymbol{e}(\boldsymbol{p},2) = \dfrac{\boldsymbol{p}}{|\boldsymbol{p}|} = \dfrac{1}{|\boldsymbol{p}|}(p_1, p_2, p_3) \end{cases}. \tag{36}$$

Obviously, as $p_3 \to |\boldsymbol{p}| = k_3$ such that $\boldsymbol{p} \to (0,0,p_3)$, one has $(\boldsymbol{e}_i)_j = \delta_{ij}$, i.e., $\boldsymbol{a}_i = \boldsymbol{e}_i$ ($i, j = 1,2,3$). For a spin-1 particle with the 3D momentum $\boldsymbol{p}$, the base vectors $\{\boldsymbol{e}_1, \boldsymbol{e}_2, \boldsymbol{e}_3\}$ can represent its three linear-polarization vectors. In fact, let $\boldsymbol{\varepsilon}(\boldsymbol{p},\lambda)$ denote the vector form of $\varepsilon(\boldsymbol{p},\lambda)$ (given by Eq. (11) with the replacements of $k_i \to p_i$, $i = 1,2,3$), one has

$$\begin{cases} \boldsymbol{\varepsilon}(\boldsymbol{p},1) = [\boldsymbol{e}(\boldsymbol{p},1) + \mathrm{i}\boldsymbol{e}(\boldsymbol{p},2)]/\sqrt{2} \\ \boldsymbol{\varepsilon}(\boldsymbol{p},-1) = [\boldsymbol{e}(\boldsymbol{p},1) - \mathrm{i}\boldsymbol{e}(\boldsymbol{p},2)]/\sqrt{2} \\ \boldsymbol{\varepsilon}(\boldsymbol{p},0) = \boldsymbol{e}(\boldsymbol{p},3) = \boldsymbol{p}/|\boldsymbol{p}| \end{cases}. \tag{37}$$

That is, $\{\boldsymbol{\varepsilon}(\boldsymbol{p},\lambda),\ \lambda = \pm 1, 0\}$ is the spinor representation of $\{\boldsymbol{e}_1, \boldsymbol{e}_2, \boldsymbol{e}_3\}$ and represents three circular polarization vectors of the spin-1 particle.

In the coordinate system $\{\boldsymbol{a}_1, \boldsymbol{a}_2, \boldsymbol{a}_3\}$, one can read $\boldsymbol{k} = \boldsymbol{k}_\perp + \boldsymbol{k}_{||}$, where

$$\boldsymbol{k}_\perp \equiv \boldsymbol{e}(\boldsymbol{p},1)k_1 + \boldsymbol{e}(\boldsymbol{p},2)k_2,\quad \boldsymbol{k}_{||} = \boldsymbol{p} = \sum\nolimits_i \boldsymbol{a}_i p_i = \boldsymbol{e}_3 k_3, \tag{38}$$

stand for $\boldsymbol{k}$'s components being perpendicular and parallel to the waveguide, respectively. Furthermore, one can write the dispersion relation of photons inside the waveguide as $E^2 = \boldsymbol{p}^2 + m^2$, it has the same form as the relativistic dispersion relation of free massive particles, where the cutoff frequency $\omega_c = m = |\boldsymbol{k}_\perp|$ plays the role of rest mass, while $\boldsymbol{p}$ represents the momentum of photons along the waveguide, such that the photons moving through the waveguide have an effective 4D momentum $p_{\mathrm{L}\mu} \equiv (E, \boldsymbol{p})$.

According to the waveguide theory, the group velocity ($\boldsymbol{v}_\mathrm{g}$) and phase velocity ($\boldsymbol{v}_\mathrm{p}$) of photons along the waveguide are, respectively (note that $\hbar = c = 1$)

$$\begin{cases} \boldsymbol{v}_{\mathrm{g}} = \boldsymbol{e}(\boldsymbol{p},3)\sqrt{1-(\omega_{\mathrm{c}}/\omega)^2} = \boldsymbol{p}/E \\ \boldsymbol{v}_{\mathrm{p}} = \boldsymbol{e}(\boldsymbol{p},3)\Big/\sqrt{1-(\omega_{\mathrm{c}}/\omega)^2} = \boldsymbol{e}(\boldsymbol{p},3)\,E/|\boldsymbol{p}| \end{cases}. \tag{39}$$

Then one can obtain the following de Broglie's relations:

$$\begin{cases} \boldsymbol{v}_{\mathrm{g}} \cdot \boldsymbol{v}_{\mathrm{p}} = c^2 = 1 \\ \boldsymbol{p} = \hbar \boldsymbol{k}_{\|} = \boldsymbol{k}_{\|} \\ E = \hbar\omega = \omega = \sqrt{m^2 + \boldsymbol{p}^2} \end{cases}. \tag{40}$$

Using $m = \omega_{\mathrm{c}}$ and Eqs. (39)-(40) one has

$$E = \frac{mc^2}{\sqrt{1-(\boldsymbol{v}_{\mathrm{g}}^2/c^2)}} = \frac{m}{\sqrt{1-\boldsymbol{v}_{\mathrm{g}}^2}}. \tag{41}$$

This is exactly the relativistic energy formula. In fact, the group velocity $\boldsymbol{v}_{\mathrm{g}}$ can be viewed as a relative velocity between an observer and a guided photon with the effective mass $m = \omega_{\mathrm{c}}$. Eqs. (39)-(41) show that the behaviors of guided waves are the same as those of de Broglie matter waves, such that the guided photon can be treated as a free massive particle. (Conversely, for a massive particle such as the electron, its zitterbewegung phenomenon shows that its motion velocity is an average one of velocity-of-light zigzag motion, just as the group velocity of the electromagnetic waves that are reflected back and forth by perfectly conducting walls as they propagate along the length of a hollow waveguide).

As we know, a light-like 4-vector can be orthogonally decomposed as the sum of a space-like 4-vector and a time-like 4-vector. In our case, the time-like part of the light-like 4-momentum $k_\mu = (\omega, \boldsymbol{k})$ is the effective 4D momentum $p_{\mathrm{L}\mu} = (E, \boldsymbol{p})$, it represents the 4D momentum of photons moving along the waveguide, and is called the traveling-wave or active part of $k_\mu$; the space-like part of $k_\mu$ is the 4D momentum $p_{\mathrm{T}\mu} \equiv (0, \boldsymbol{k}_\perp) = m\eta_\mu$ ($\eta_\mu \equiv (0, \boldsymbol{k}_\perp/m)$ satisfies $\eta_\mu \eta^\mu = -1$), it contributes the effective mass, and is called the stationary-wave or frozen part of $k_\mu$. In other words, as the *rest* energy of photons inside the waveguide (i.e., the energy as the group velocity $\boldsymbol{v}_{\mathrm{g}}=0$), the effective mass arises by freezing out the degrees of freedom related to the transverse motion of photons inside the waveguide. Therefore, we obtain an orthogonal decomposition for $k_\mu=(\omega,\boldsymbol{k})$ as follows:

$$k_\mu=(\omega,\boldsymbol{k})=p_{\mathrm{T}\mu}+p_{\mathrm{L}\mu},\ \ p_{\mathrm{T}\mu}\equiv(0,\boldsymbol{k}_\perp)=m\eta_\mu,\ \ p_{\mathrm{L}\mu}=(E,\boldsymbol{p}). \tag{42}$$

Such an orthogonal decomposition is Lorentz invariant because of $p_{\mathrm{L}\mu}p_{\mathrm{T}}^{\mu}=0$.

Likewise, as for $x_\mu=(t,\boldsymbol{x})$ with $\boldsymbol{x}=\sum_i \boldsymbol{e}_i x_i=(x_1,x_2,x_3)$, in the coordinate system $\{\boldsymbol{a}_1,\boldsymbol{a}_2,\boldsymbol{a}_3\}$ one has $\boldsymbol{e}_3x_3=\sum_j \boldsymbol{a}_j r_j=\boldsymbol{r}$, i.e., the 3D vector $\boldsymbol{r}$ is parallel to the waveguide. Let $\boldsymbol{x}_\perp\equiv\boldsymbol{e}_1x_1+\boldsymbol{e}_2x_2$, the orthogonal decomposition for $x_\mu$ can be written as

$$x_\mu=(t,\boldsymbol{x})=x_{\mathrm{T}\mu}+x_{\mathrm{L}\mu},\ \ x_{\mathrm{T}\mu}\equiv(0,\boldsymbol{x}_\perp),\ \ x_{\mathrm{L}\mu}=(t,\boldsymbol{r}). \tag{43}$$

It is easy to show that

$$k_\mu x^\mu=(p_{\mathrm{T}\mu}+p_{\mathrm{L}\mu})(x_{\mathrm{T}}^{\mu}+x_{\mathrm{L}}^{\mu})=p_{\mathrm{T}\mu}x_{\mathrm{T}}^{\mu}+p_{\mathrm{L}\mu}x_{\mathrm{L}}^{\mu}. \tag{44}$$

The operator $\hat{p}_\mu=\mathrm{i}\partial_\mu=\mathrm{i}\,\partial/\partial x^\mu$ represents the totally 4D momentum operator of photons inside the waveguide, while

$$\hat{p}_{\mathrm{L}\mu}=\mathrm{i}\partial_{\mathrm{L}\mu}=\mathrm{i}\,\partial/\partial x_{\mathrm{L}}^{\mu}, \tag{45}$$

represents the 4D momentum operator of photons moving along the waveguide. In spite of the boundary conditions for the waveguide, there are no charges in the free space inside the waveguide, and then where photons should obey the Dirac-like equation $\mathrm{i}\beta^\mu\partial_\mu\psi(x)=0$. Because of $\psi(x)\sim\exp(-\mathrm{i}k_\mu x^\mu)=\exp[-\mathrm{i}(p_{\mathrm{T}\mu}x_{\mathrm{T}}^{\mu}+p_{\mathrm{L}\mu}x_{\mathrm{L}}^{\mu})]$, one has $p_{\mathrm{L}\mu}\psi(x)=\mathrm{i}\partial_{\mathrm{L}\mu}\psi(x)$, and then

$$\mathrm{i}\beta^\mu\partial_\mu\psi(x)=\beta^\mu k_\mu\psi(x)=\beta^\mu(p_{\mathrm{L}\mu}+p_{\mathrm{T}\mu})\psi(x)=\beta^\mu(\mathrm{i}\partial_{\mathrm{L}\mu}+p_{\mathrm{T}\mu})\psi(x). \tag{46}$$

Using $\mathrm{i}\beta^\mu\partial_\mu\psi(x)=0$ and $p_{\mathrm{T}\mu}=m\eta_\mu$, from Eq. (46) one has $\mathrm{i}\beta^\mu(\partial_{\mathrm{L}\mu}-\mathrm{i}m\eta_\mu)\psi(x)=0$.

Let $\psi(x)=\varphi(x_{\mathrm{L}})\exp(-\mathrm{i}p_{\mathrm{T}\mu}x_{\mathrm{T}}^{\mu})$, obviously $\varphi(x_{\mathrm{L}})\sim\exp(-\mathrm{i}p_{\mathrm{L}\mu}x_{\mathrm{L}}^{\mu})$, we obtain the Dirac-like equation of photons moving along the waveguide

$$\mathrm{i}\beta^{\mu}(\partial_{\mathrm{L}\mu}-\mathrm{i}m\eta_{\mu})\varphi(x_{\mathrm{L}})=0\,. \tag{47}$$

Eq. (47) is called the Dirac-like equation of photons with the effective mass *m*. Using Eq. (47) and $\eta^{\mu}\partial_{\mathrm{L}\mu}\varphi(x_{\mathrm{L}})=\partial_{\mathrm{L}\mu}\eta^{\mu}\varphi(x_{\mathrm{L}})=0$ one can obtain the Klein-Gordon equation

$$(\partial_{\mathrm{L}\mu}\partial_{\mathrm{L}}^{\mu}+m^{2})\varphi(x_{\mathrm{L}})=0\,. \tag{48}$$

In the first-quantized sense, Eqs. (47)-(48) serve as the relativistic quantum-mechanical equations of photons moving along the waveguide, while in the second-quantized sense, they are quantum-field-theory equations. It is important to note that, Eqs. (47)-(48) are covariant under Lorentz boosts along the direction of the waveguide, and the effective mass *m* is an invariant quantity under such Lorentz transformations. The effective mass *m* of photons inside the waveguide is produced via the symmetry breaking from SO (1, 3) to SO (1, 1). From another point of view, only those modes in the form of transverse electric (TE) and transverse magnetic (TM) modes can propagate in the waveguide, that is, electromagnetic waves inside the waveguide possess the degree of freedom of longitudinal polarization and in such a way they obtain the effective mass *m*. In addition to the Higgs mechanism, here we might provide another heuristic way of introducing the vector field's mass (maybe the origin of mass is not unique).

In terms of the effective mass $m=\omega_{\mathrm{c}}$, one can define the equivalent Compton wavelength of photons moving along the waveguide ($\hbar=c=1$)

$$\lambda\!\!\!^{-}_{\mathrm{c}}\equiv\hbar c/\hbar\omega_{\mathrm{c}}=1/m\,. \tag{49}$$

As we know, it is impossible to localize a massive particle with a greater precision than its Compton wavelength, which owing to many-particle phenomena. Similarly, it is impossible to localize a photon inside a waveguide with a greater precision than its equivalent Compton wavelength, which owing to evanescent-wave phenomena (note that for a travelling wave inside the waveguide, there is always an inertial reference frame in which its group velocity vanishes).

## IV. SPACE-LIKE BEHAVIORS OF EVANESCENT GUIDED WAVES

For convenience, from now on, the 4D space-time coordinate of photons moving along the waveguide is rewritten as $x_\mu = (t, \boldsymbol{x})$ instead of the original $x_{\mathrm{L}\mu} = (t, \boldsymbol{r})$, and then $\hat{p}_{\mathrm{L}\mu} = \mathrm{i}\partial_{\mathrm{L}\mu} = \mathrm{i}\,\partial/\partial x_{\mathrm{L}}^{\mu}$ is rewritten as $\hat{p}_\mu = \mathrm{i}\partial_\mu = \mathrm{i}\,\partial/\partial x^\mu$. A free Feynman propagator (say, $S(x-y)$) represents the probability amplitude for a particle to propagate from $y$ to $x$. Therefore, to study the space-like behaviors of photons inside an undersized waveguide, we will analyze the free Feynman propagator of photons along the waveguide. Applying the discussions in **Section III** and the definition Eq. (31), one can obtain the free Feynman propagator of photons with the effective mass *m* as follows (here we rewrite $\mathrm{i}R_{\mathrm{f}}(x_1 - x_2)$ as $\mathrm{i}G_{\mathrm{f}}(x_1 - x_2)$):

$$\mathrm{i}G_{\mathrm{f}}(x_1 - x_2) = \mathrm{i}\beta^\mu(\partial_\mu - \mathrm{i}m\eta_\mu)\delta'_{\mathrm{T}}\mathrm{i}S(x_1 - x_2)\,. \tag{50}$$

where the new transverse delta-function $\delta'_{\mathrm{T}}$ has component representation ($i, j = 1, 2, 3$):

$$\delta'_{\mathrm{T}ij} = \delta_{ij} - [(\nabla_i + \mathrm{i}m\eta_i)(\nabla_j + \mathrm{i}m\eta_j)/(\nabla + \mathrm{i}m\boldsymbol{\eta})^2]\,. \tag{51}$$

The function $S(x)$ presented in Eq. (50) can be expressed as:

$$S(x) = \begin{cases} S_1(x) = -\mathrm{i}\displaystyle\int_{-\infty}^{+\infty} \frac{\mathrm{d}^3 p}{(2\pi)^3}\frac{1}{2E}\exp(-\mathrm{i}E|x_0| + \mathrm{i}\boldsymbol{p}\cdot\boldsymbol{x}), & \text{for } E = \sqrt{m^2 + \boldsymbol{p}^2} \geq m \\ S_2(x) = -\displaystyle\int_{-m}^{+m} \frac{\mathrm{d}^3 q}{(2\pi)^3}\frac{1}{2E}\exp(-\mathrm{i}E|x_0| - \boldsymbol{q}\cdot\boldsymbol{x}), & \text{for } 0 < E = \sqrt{m^2 - \boldsymbol{q}^2} < m \end{cases}, \tag{52}$$

where the function $S_1(x)$ is the free Feynman propagator of the Klein-Gordon field. In fact,

for $E \geq m$ one can prove that $\mathrm{i}G_{\mathrm{f}}(x_1 - x_2)$ is the Green function of Eq. (47).

$$\mathrm{i}\beta^{\mu}(\partial_{\mu} - \mathrm{i}m\eta_{\mu})G_{\mathrm{f}}(x_1 - x_2) = \delta^4(x_1 - x_2). \tag{53}$$

Eq. (50) implies that the space-like behaviors of photons with the effective mass $m$ can be displayed via the function $S(x)$. In fact, all massive fields satisfy the Klein-Gordon equation (48) such that their free Feynman propagators are all related to the function $S_1(x)$, and their space-like behaviors are discussed usually by analyzing the space-like behaviors of $S_1(x)$. Likewise, our starting point is to analyze the space-like behaviors of the function $S(x)$. However, in contrary to the usual quantum field theory, our discussion also includes the case of $0 < E < m$.

It is very important to note that, $m = \omega_{\mathrm{c}} = \pi/b_1$ is the lowest-order cutoff frequency, and Eqs. (47)-(48) are written in an arbitrary inertial frame of reference. Therefore, the guided waves with $E = \omega \geq m = \omega_{\mathrm{c}}$ need not be the traveling waves (but the ones with $0 < E < m$ must be the evanescent waves). On the other hand, because the evanescent waves oscillate with time as $\exp(-\mathrm{i}\omega t)$, as observed in an inertial frame of reference moving relative to the waveguide, they can propagate in a time-like way, which is related to the Lorentz transformation of $\omega t$. However, the propagation of the evanescent waves through the waveguide can be characterized by an exponential damping factor. In the following, we will show that such a damping propagation is actually a space-like one.

In order to evaluate the integral Eq. (52), let $|x_0| = t$, $|\boldsymbol{x}| = r$, and then $x^2 = x_{\mu}x^{\mu} = t^2 - r^2$. For time-like interval $x^2 > 0$, let $t = \sqrt{x^2}\cosh\phi$ and $r = \sqrt{x^2}\sinh\phi$, and there is always an inertial frame in which $r = 0$; for space-like interval $x^2 < 0$, let $t = \sqrt{-x^2}\sinh\phi$ and $r = \sqrt{-x^2}\cosh\phi$, and there is always an inertial frame in which $t = 0$. As $\phi$ varys in $[0, +\infty)$, $x^2$ is Lorentz invariant. For convenience we will take $\phi \to 0$.

Furthermore, the integral representation of the Hankel function of the second kind is useful:

$$H_0^{(2)}(z)=\frac{\mathrm{i}}{\pi}\int_{-\infty}^{+\infty}\mathrm{d}\varphi\exp(-\mathrm{i}z\cosh\varphi)=\frac{2}{\pi}\int_0^{\pi/2}\mathrm{d}\theta\exp(-\mathrm{i}z\sin\theta)\,. \tag{54}$$

Now, we separately discuss the two possible cases:

1). As $E=\omega\geq m=\omega_{\mathrm{c}}$, form Eq. (52) one can obtain (a similar reference see Ref. [19])

$$S(x)=S_1(x)=\begin{cases}\dfrac{\omega_{\mathrm{c}}}{8\pi\sqrt{x^2}}H_1^{(2)}(\omega_{\mathrm{c}}\sqrt{x^2}) & \text{for timelike separate } x^2>0\\ \dfrac{\mathrm{i}\omega_{\mathrm{c}}}{8\pi\sqrt{-x^2}}H_1^{(2)}(-\mathrm{i}\omega_{\mathrm{c}}\sqrt{-x^2}) & \text{for spacelike separate } x^2<0\\ -\dfrac{1}{4\pi}\delta(x^2) & \text{for lightlike separate } x^2=0\end{cases}, \tag{55}$$

where $H_1^{(2)}(z)=-\mathrm{d}H_0^{(2)}(z)/\mathrm{d}z$. As an important special case of $S_1(x)$, in the limit of $m=\omega_{\mathrm{c}}\to 0$, one has ($\rho\to 0$):

$$D(x)\equiv\lim_{\omega_{\mathrm{c}}\to 0}S_1(x)=\frac{\mathrm{i}}{4\pi}\frac{1}{x^2-\mathrm{i}\rho}\,. \tag{56}$$

This is related to the free-space photon propagator. In fact, in the limit of $m=\omega_{\mathrm{c}}\to 0$, the transversal dimensions of the waveguide approach to infinity such that the space inside the waveguide becomes a free space. The Hankel function behaves for large arguments $|z|$ as

$$H_\nu^{(2)}(z)\sim\sqrt{2/\pi z}\exp[-\mathrm{i}(z-\pi\nu/2-\pi/4)]\,,\quad |z|\to+\infty\,. \tag{57}$$

Applying Eq. (57) to Eq. (55) one can deduce the asymptotic behaviors of $S_1(x)$:

$$S(x)=S_1(x)\sim\begin{cases}(x^2)^{-3/4}\exp(-\mathrm{i}\omega_{\mathrm{c}}\sqrt{x^2}) & \text{for } x^2\to+\infty\\ (-x^2)^{-3/4}\exp(-\omega_{\mathrm{c}}\sqrt{-x^2}) & \text{for } x^2\to-\infty\end{cases}. \tag{58}$$

2). As $0<E<m=\omega_{\mathrm{c}}$, applying spherical polar coordinates to Eq. (52) one can obtain:

$$S(x)=S_2(x)=-\frac{1}{8\pi^2 r}\frac{\partial}{\partial r}\int_0^m\mathrm{d}\kappa\frac{1}{\sqrt{m^2-\kappa^2}}\exp(-\mathrm{i}t\sqrt{m^2-\kappa^2}-\kappa r)\,, \tag{59}$$

where $\kappa=|\boldsymbol{q}|$. We only concern the evanescent waves, and then the integrating range in Eq.

(59) is retaken as $[0,m]$, i.e., there does not contain the contribution from the anti-evanescent waves. From Eq. (59) one can obtain

$$S(x)=S_2(x)=\begin{cases}\dfrac{-\omega_{\mathrm{c}}}{16\pi\sqrt{x^2}}H_1^{(2)}(\omega_{\mathrm{c}}\sqrt{x^2}) & \text{for timelike separate } x^2>0\\ \dfrac{-\mathrm{i}\omega_{\mathrm{c}}}{16\pi\sqrt{-x^2}}H_1^{(2)}(-\mathrm{i}\omega_{\mathrm{c}}\sqrt{-x^2}) & \text{for spacelike separate } x^2<0\end{cases}. \tag{60}$$

Here for convenience the case of $x^2=0$ is not concerned, it is not our purpose. Because $S_2(x)=-S_1(x)/2$ for $x^2\neq 0$, in the limit of $|x^2|\to+\infty$, the asymptotic behaviors of $S_2(x)$ are the same as those of $S_1(x)$.

On one hand, one has the recurrence relations $z^{-\nu}H_{\nu+1}^{(2)}(z)=-\mathrm{d}[z^{-\nu}H_{\nu}^{(2)}(z)]/\mathrm{d}z$; on the other hand, Eq. (57) shows that the asymptotic behaviors of the Hankel functions $H_{\nu}^{(2)}(z)$ are similar for different order $\nu=0,1,2\ldots$ Therefore, the time-like and space-like behaviors of the propagator $\mathrm{i}G_{\mathrm{f}}(x_1-x_2)$ (given by Eq. (50)) are similar to those of the propagator $S(x)$.

Furthermore, all physical conclusions here would be preserved if our discussion were performed in the coordinate system $\{\boldsymbol{e}_1,\boldsymbol{e}_2,\boldsymbol{e}_3\}$ (in which the waveguide is placed along the direction of $\boldsymbol{e}_3$ such that $\boldsymbol{p}=(0,0,p_3)$), for the moment our discussion would reduction to a (1+1) D issue, and $S(x)$ is related to $H_0^{(2)}(z)$ rather than $H_1^{(2)}(z)$. In fact, for a given waveguide, the momentum direction of photons propagating along the waveguide is fixed, though Eqs. (47)-(48) are written in the 4D form, the integral measures in Eq. (52) should be $\mathrm{d}p/2\pi$ and $\mathrm{d}q/2\pi$ ($\boldsymbol{p}$ and $\boldsymbol{q}$ are parallel to the waveguide), such that $S(x)$ is related to $H_0^{(2)}(z)$ rather than $H_1^{(2)}(z)$, but this does not alter any physical conclusion. To compare our results with those in traditional quantum field theory, the integral measures in Eq. (52) are taken as $\mathrm{d}^3p/(2\pi)^3$ and $\mathrm{d}^3q/(2\pi)^3$.

To sum up, Eqs. (55), (58) and (60) together show that, for time-like distances ($x^2 > 0$) the propagator is an oscillating function slowly decreasing in amplitude owing to the power-law factor; for space-like distances ($x^2 < 0$) the propagator rapidly falls to zero according to the exponential function, where the scale is set by the inverse cut-off frequency of the waveguide (i.e., the equivalent Compton wavelength of photons inside the waveguide). As mentioned before, the propagation of the evanescent waves through an undersized waveguide is characterized by an exponential damping factor. Therefore, the fact that the propagator can be nonzero (albeit rapidly decreasing) also in the region of space-like distances, corresponds to the fact that the evanescent waves can propagate through the waveguide (i.e., the photonic tunneling phenomenon), which is caused by the difficulty to localize a photon on a scale smaller than its equivalent Compton wavelength. In fact, a similar statement can be found in standard quantum field theory [19].

By the way, if the evanescent and anti-evanescent waves are considered together, i.e., the integrating range in Eq. (59) is taken as $[-m, m]$, one has

$$S_2(x) = \begin{cases} -S_1(x) & \text{for timelike separate } x^2 > 0 \\ \dfrac{\omega_c}{8\pi\sqrt{-x^2}} I_1(\omega_c\sqrt{-x^2}) & \text{for spacelike separate } x^2 < 0 \end{cases}, \tag{61}$$

where $I_1(z)$ is the modified Bessel function of the first kind. For the moment, for large space-like distances ($x^2 \to -\infty$) one has $S_2(x) \to \exp(\omega_c\sqrt{-x^2})$, its physical meaning remains to be investigated. In terms of a normalization probability $P$ (being proportional to a space integral of $|S_2(x)|^2$), one has $P = 0$ for $x^2 > 0$, while $P = 1$ for $x^2 < 0$, which implies that the space-like behavior will occur to a certainty.

## V. QUANTUM-MECHANICAL CAUSALITY

As discussed above, the propagation of the evanescent waves through an undersized waveguide is a superluminal one. This space-like process is due to a purely quantum-mechanical effect, and its causality can be discussed from the following three aspects (they are incident with each other):

1). To avoid a possible causality paradox, one can resort to the particle-antiparticle symmetry. The process of a particle created at x and annihilated at y as observed in a frame of reference, is identical with that of an antiparticle created at y and annihilated at x as observed in another frame of reference [20]. In our case, the antiparticle of the photon is the photon itself. Therefore, the process that a photon propagates superluminally from A to B as observed in a frame of reference, is equivalent to that the photon propagates superluminally from B to A as observed in another frame of reference.

2). Though a particle can propagate over a space-like interval, causality is preserved provided that a measurement performed at one point cannot affect another measurement at a point separated from the first with a space-like interval [21], this implies that the commutator between two observables taken at two space-like separated points has to vanish. However, it is very important to note that, the causality condition mentioned in some quantum-field-theory textbooks (e.g., Ref. [21]), i.e., that "the commutator between two field operators located at space-like distance must vanish", is improper provided that these field operators are not observable quantities [22-23]. In our case, the canonical commutation relations given by Eq. (25) with $t=t'$ and $\boldsymbol{x}\neq\boldsymbol{x}'$, vanish for two propagation modes, but do not vanish for two evanescent modes (in agreement with the conclusion presented in Ref. [24]). Because evanescent modes inside an undersized waveguide are not observable [17], causality is preserved. A more detailed discussion can be found in Ref. [25].

3). It is useful to distinguish between two notions of causality [26]: a) strong causality. By this we mean that for each individual experiment in which two systems, separated by a distance $R$, are prepared at time $t$=0, no disturbance or excitation of the second system occurs for $t<R/c$. In Ref. [26] the author shows that strong causality cannot be checked or it may fail in a theory. b) weak causality. It means causality for expectation values or ensemble average only, not for individual process. In a strict sense, weak causality can only be checked experimentally for infinite ensembles. Within local quantum field theory a rigorous proof of weak causality for local observables has been given in the previous literatures [27-28]. In our case, weak causality is valid.

## VI. CONCLUSIONS

The photon field inside a waveguide can be described by the Dirac-like equation of photons with the effective mass $m$ (see Eq. (47)), and one can investigate the propagation of the evanescent waves through an undersized waveguide in virtue of photon's quantum theory itself, rather than via a quantum-mechanical analogy. In particular, we show that the superluminal propagation of the guided evanescent waves corresponds to a known conclusion in quantum field theory and preserves causality in a quantum-mechanical manner. In the previous theoretical investigations, the conclusion that the propagation of the evanescent waves is a superluminal one is based on the theory of tunneling time. However, there are a lot of controversies about the theoretical model of tunneling time [29-32]. As a consequence, some people do not think that the guided evanescent waves possess superluminal behavior indeed [33-35]. In contrast to this, our conclusion is obtained within the framework of quantum field theory and is rigorous.

**ACKNOWLEDGMENTS**

The first author (Z. Y. Wang) would like to greatly thank Prof. G. Nimtz for his helpful discussions. Project supported by the Specialized Research Fund for the Doctoral Program of Higher Education of China (Grant No. 20050614022) and by the National Natural Science Foundation of China (Grant No. 60671030).

---

*E-Mail: zywang@uestc.edu.cn

[1] M. Campi and M. Harrison, Am. J. Phys. **35**, 133 (1967).

[2] A. Enders and G. Nimtz, Phys. Rev. E **48**, 632 (1993).

[3] G. Nimtz and W. Heitmann, Progr. Quantum Electron. **21**, 81 (1997).

[4] T. D. Newton and E. P. Wigner, Rev. Mod. Phys. **21**, 400 (1949).

[5] A. S. Wightman, Rev. Mod. Phys. **34**, 854 (1962).

[6] L. Mandel, Phys. Rev. **144**, 1071 (1966).

[7] R. J. Cook, Phys. Rev. A **25**, 2164(1982); ibid, **26**, 2754 (1982).

[8] T. Inagaki, Phys. Rev. A **49**, 2839 (1994); ibid, **57**, 2204 (1998).

[9] I. Bialynicki-Birula, Acta Phys. Polon. A **86**, 97 (1994).

[10] I. Bialynicki-Birula, *Progress in Optics XXXVI*, edited by E. Wolf (Elsevier, Amsterdam, 1996).

[11] I. Bialynicki-Birula, *Coherence and Quantum Optics* VII, edited by J. Eberly, L. Mandel, and E. Wolf (Plenum Press, New York, 1996), p. 313

[12] J. E. Sipe, Phys. Rev. A **52**, 1875 (1994).

[13] O. Keller, Phys. Rep. **411**, 1 (2005).

[14] C. Adlard, E. R. Pike, and S. Sarkar, Phys. Rev. Lett **79**, 1585 (1997).

[15] I. Bialynicki-Birula, Phys. Rev. Lett **80**, 5247 (1998).

[16] K. W. Chan, C. K. Law, and J. H. Eberly, Phys. Rev. Lett **88**, 100402 (2002).

[17] A.A. Stahlhofen, G. Nimtz, Europhys. Lett. **76**, 189 (2006).

[18] J. D. Jackson, *Classical Electrodynamics* (2nd), John Wiley & Sonc inc., New York, 1975, pp744-752 and pp391-398.

[19] W. Greiner and J. Reinhardt, *Quantum Electrodynamics*, Springer-Verlag Berlin Heidelberg New York, 1992, pp. 55-60.

[20] S. Weinberg, *Gravitation and Cosmology*, John Wiley, New York 1972, Section 2-13.

[21] M. E. Peskin and D. V. Schroeder, *An Introduction to Quantum Field Theory*, Addison-Wesley Publishing Company, New York , 1995, pp. 27-33.

[22] W. Greiner, J. Reinhardt, *FIELD QUANTIZATION* (Springer-Verlag Berlin Heidelberg, 1996), pp.208-209

[23] S. Weinberg, *The Quantum Theory of Fields* (Vol.1) (Cambridge University Press, England 1995), pp.198-199.

[24] C. K. Carniglia and L. Mandel, Phys. Rev. D **1**, 280 (1971).

[25] G. Nimtz, Lecture Notes in Physics (Springer), Vol. 702, 509 (2006).

[26] G. C. Hegerfeldt, Ann. Phys. (Leipzig) **7**, 716 (1998).

[27] D. Buchholz and J. Yngvason, Phys. Rev. Lett. **73**, 613 (1994).

[28] H. Neumann and R. Werner, Intern. J. Theor. Phys. **22**, 781 (1983).

[29] E. H. Hauge and J. A. Støvneng, Rev. Mod. Phys. **61**, 917 (1989).

[30] R. Landauer and Th. Martin, Rev. Mod. Phys. **66**, 217 (1994).

[31] J. G. Muga, R. Sala, and I. L. Egusguiza, *Time in Quantum Mechanics* (Springer, Verlag Berlin Heidelberg, 2002).

[32] V. S. Olkhovsky, E. Recami and J. Jakiel, Phys. Reports **398**, 133 (2004).

[33] A. D. Jackson, A. Lande and B. Lautrup, Phys. Rev. A **64**, 044101 (2001).

[34] H. G. Winful, Phys. Rev. E **68**, 016615 (2003).

[35] D. Sokolovski, A. Z. Msezane and V. R. Shaginyan, Phys. Rev. A **71**, 064103 (2005).